\documentclass[prd,aps,onecolumn,preprintnumbers, showpacs,
nofootinbib,superscriptaddress,notitlepage]{revtex4-1}
\usepackage[english]{babel}
\usepackage{verbatim}
\usepackage{xcolor}

\usepackage[normalem]{ulem}

\usepackage{amsmath}
\usepackage{graphicx}
\usepackage[colorlinks=true, allcolors=blue]{hyperref}
\begin{document}

\title{Lattice QCD Calculations of Parton Physics}
\author{Martha Constantinou}\affiliation{Department of Physics, Temple University, 1925 N. 12th Street, Philadelphia, PA 19122-1801, USA}
\author{Luigi Del Debbio}\affiliation{Higgs Centre for Theoretical Physics, University of Edinburgh, Peter Guthrie Tait Road, Edinburgh EH9 3FD, United Kingdom}
\author{Xiangdong Ji}\affiliation{Department of Physics, University of Maryland, College Park, MD 20742, USA}
\author{Huey-Wen Lin}
\affiliation{Department of Physics and Astronomy, Michigan State University, East Lansing, MI 48824}
\affiliation{Department of Computational Mathematics, Science \& Engineering, Michigan State University, East Lansing, MI 48824}
\author{Keh-Fei Liu}\affiliation{Department of Physics and Astronomy, University of Kentucky, Lexington, KY 40506, USA}
\author{Christopher Monahan}\affiliation{Department of Physics, William \& Mary, Williamsburg, VA 23187, USA}
\affiliation{Thomas Jefferson National Accelerator Facility,  Newport News, VA 23606, USA}
\author{Kostas Orginos}\affiliation{Department of Physics, William \& Mary, Williamsburg, VA 23187, USA}
\affiliation{Thomas Jefferson National Accelerator Facility,  Newport News, VA 23606, USA}
\author{Peter Petreczky}\affiliation{Physics Department, Brookhaven National Laboratory, Upton, New York 11973, USA }
\author{Jian-Wei Qiu}\affiliation{Thomas Jefferson National Accelerator Facility,  Newport News, VA 23606, USA}
\author{David Richards}\affiliation{Thomas Jefferson National Accelerator Facility,  Newport News, VA 23606, USA}
\author{Nobuo Sato}\affiliation{Thomas Jefferson National Accelerator Facility,  Newport News, VA 23606, USA}
\author{Phiala Shanahan}\affiliation{Center for Theoretical Physics, Massachusetts Institute of Technology, Cambridge, MA 02139, USA}
\author{C.-P. Yuan}
\affiliation{Department of Physics and Astronomy, Michigan State University, East Lansing, MI 48824}
\author{Jian-Hui Zhang}\affiliation{Center of Advanced Quantum Studies, Department of Physics, Beijing Normal University, Beijing 100875, China }
\author{Yong Zhao}\affiliation{Physics Division, Argonne National Laboratory, Lemont, IL 60439, USA}

\begin{abstract}
    In this document, we summarize the status and challenges of calculating parton physics in lattice QCD for the US Particle Physics Community Planning Exercise (a.k.a. ``Snowmass''). While PDF-moments calculations have been very successful and been continuously improved, new methods
    have been developed to calculate distributions directly in $x$-space. Many recent lattice studies have been focused on calculating isovector PDFs of the pion and nucleon, learning to control systematics associated with excited-state contamination, renormalization and continuum extrapolations, pion-mass and finite-volume effects, etc. Although in some cases, the lattice results are already competitive with experimental data, to reach the level of precision in a wide range of $x$ for unpolarized nucleon PDFs impactful for future
    collider physics remains a challenge, and may require exascale supercomputing power. The new theoretical methods open the door for calculating other
    partonic observables which will be the focus of the experimental program in nuclear physics, including generalized parton distributions and transverse-momentum dependent PDFs. A fruitful interplay between experimental data and lattice-QCD calculations
    will usher in a new era for parton physics and hadron structure.
    
\end{abstract}
\date{\today}
 
\preprint{JLAB-THY-22-3564,MIT-CTP/5408,MSUHEP-22-004}

\maketitle

\section{Introduction and Executive Summary}

Feynman's parton has become one of the most fundamental
concepts for describing high-energy scattering of hadrons at colliders searching for new physics beyond the standard model of particle physics. It has also provided a basic language to describe the femto-structure of hadrons such as nucleons, pions, and heavy-flavored states from the underlying theory of quarks and gluons: quantum chromodynamics (QCD). Calculating the partonic
structure of QCD bound states from the first principles with controlled accuracy remains an important unsolved problem in the standard model. This white paper 
summarizes the status of efforts and challenges in lattice QCD calculations of parton physics and the interplay of lattice calculations with phenomenological determinations of partonic structures from experimental data, and
highlights some future prospects for progress in 
the next five to ten years.

Precision determination of PDFs is not only important for probing the standard model, but also to advance interpretation of high-energy experiments searching for signs of physics beyond the standard model.
In addition to energy-frontier experiments like the LHC, there are also many mid-energy experimental efforts around the world, such as at Brookhaven and Jefferson Laboratories in the United States,  GSI in Germany, J-PARC in Japan, or a future EIC.
These are set to explore the less-known kinematic regions of nucleon structure and more.
The pursuit of PDFs has led to collaborations of theorists and experimentalists working side-by-side to take advantage of all available data, evaluating different combinations of input theories, parameter choices and assumptions, resulting in multiple global-PDF determinations (see references within Refs.~\cite{Constantinou:2020hdm,Lin:2017snn} for example ).

The simplest quantities to calculate on the lattice are moments of the collinear parton distribution functions (PDFs), which provide momentum-space ``global'' information about partons. It is, however, not easy to connect PDFs directly to a particular experiment in which particles of definite momentum are measured. A more desirable theoretical approach is to access directly the $x$-dependence or ``local'' information in momentum space. Broadly speaking, two approaches have been developed for lattice QCD in recent years. The first focuses on the 
short-distance factorization (SDF) in coordinate space, and the other is based on an expansion in terms of a large hadron momentum, large-momentum effective theory (LaMET). Both methods require calculating coordinate-space correlation functions in large-momentum hadron states. 
Extensive studies in the last few years have been made to validate these approaches, and much has been learned about Monte-Carlo data generation and analysis, and various types of systematic uncertainties to be controlled. Meanwhile, much lies ahead to bring these methods to bear on high-precision calculations applicable for particle physics at high-energy colliders, and on theoretical predictions for the parton structure of hadrons to be studied at Jefferson Lab 12~GeV and the future Electron-Ion Collider (EIC) at Brookhaven National Lab. 

Taking stock of the current status of lattice calculations of parton physics, we find: 

\begin{itemize}
\item 
Extensive development in methods has been made for calculating collinear PDFs, generalized parton distributions (GPDs), TMD distributions and evolution kernels. While one-loop matching kernels are widely available, high-precision calculations require two-loop (only available for isovector PDFs), or higher-order matching, as well as quantitative understanding of renormalon uncertainties and higher-twist effects. Methods need to be developed to overcome challenges in systematic uncertainties arising from inverse problems and coordinate-space extrapolations at large distances. 

\item  While many lattice exploratory studies have been undertaken, a large quantity of high-statistics lattice-QCD data spanning different hadron momenta, quark masses, lattice spacing, volumes, valence and sea quarks, are needed for systematic analysis. New methods are needed to increase the signal-to-noise ratio for hadronic matrix elements, particularly for large hadron momentum and large spatial correlations. Criteria need to be established to reduce the excited-state contamination, finite-volume effects, and the effects of nonzero lattice spacing. 

\item 
Extensive calculations have been carried out for isovector PDFs and distribution amplitudes (DAs). Precision calculations requiring control over systematics from renormalization, continuum limit, 
inverse problem in SDF and extrapolation to large lightcone distance in LaMET. Calibrations can be made against lattice moments and high-precision experimental data. 
Closure tests with artificial data can be used to assess the robustness of the current procedures. 
Development calculation of the 5\% level (total systematics) isovector or better precision in collinear PDFs and 
improvement the precision of the current PDF calculations including sea quark distributions (as well as large $x$ quarks and gluons) will require significant increase in computational resources to make an impact in high-energy collider experiments. 

\item 
Exploratory calculations have been undertaken for quantities beyond the twist-2 collinear PDFs, including GPDs, higher-twist PDFs, and new TMD-dependent quantities, such as the Collins-Soper rapidity evolution kernel, and soft functions. Systematic lattice calculations of these observables will be crucial for the experimental programs at JLab and the EIC.

\item Lattice matrix elements are complementary to experimental data, and in certain cases they can be used together to generate hybrid state-of-art parton distribution sets~\cite{Lin:2017stx,Cichy:2019ebf,Bringewatt:2020ixn,DelDebbio:2020rgv}. This is particularly important 
for three-dimensional nuclear femtography, because extracting the GPDs from experimental data alone can be extremely challenging. The connected-sea and disconnected-sea partons are innately coded in lattice calculations of the PDF, GPD and TMD via the connected insertion and disconnected insertion. Such lattice calculations of the distributions at small $x$ and the moments are needed to help the global fitting of the connected-sea and disconnected-sea from the experimental results. This separation is essential in understanding the origin of the ``proton spin crisis'' for example. 
\end{itemize}

We emphasize that just like published experimental data, it is important to standardize and share the raw lattice QCD data from different collaborations to test theoretical ideas and benchmark different calculations and methods.
In the remainder of the white paper, we explain the above findings in detail. 

\section{Lattice Parton Methods} \label{parton_methods}

Collinear PDFs have been traditionally studied on the lattice in terms of the matrix
elements of local twist-2 operators, e.g.,~\cite{Peskin:1995ev}
\begin{equation}
    O^{\mu_1\mu_2...\mu_n}_q = 
    \overline{\psi}_q \gamma^{(\mu_1}iD^{\mu_2}...
    iD^{\mu_n)}\psi_q \ , 
\end{equation}
where all Lorentz indices among $(\cdots)$ are symmetrized and trace-subtracted, and the hadron states can have zero momentum, $\vec{P}=0$. The above matrix elements yield the Mellin $x$-moments of PDFs. This approach follows from the QCD factorization theorems in which PDFs emerge as Fourier transformations of light-cone correlations of quark and gluon fields in hadrons, 
\begin{equation}
     f_q (x) = \frac{1}{2}
     \int \frac{d\lambda}{2\pi}
     e^{ix\lambda} \langle P|\overline{\psi}_q(0)W(0,\lambda n) 
     \!\! \not\! n \psi_q(\lambda n)|P\rangle \ , 
\end{equation}
where $n^\mu=0$ is a light-like four vector, $n^2=0$, and $W$ is a light-cone gauge link. For a thorough review and discussions on the achievements using this approach, see Refs.~\cite{Lin:2017snn,Constantinou:2020hdm}. The overall results support the expectation that the parton structure of hadrons can ultimately be explained in terms of fundamental QCD. The limitations and possible solutions for moment approaches have been studied by various groups over the years~\cite{Davoudi:2012ya}. 

\subsection{Current-current Correlator and Short-Distance Factorization for Collinear PDFs}

To go beyond individual moments, the deep-inelastic scattering (DIS) hadron tensor and related Compton amplitude, and more generally current-current correlators in Euclidean space,
\begin{equation}
       \langle P |J^\mu(0)J^\nu(z)|P\rangle 
\end{equation}
where $z$ is a Euclidean four-coordinate, have been exploited to generate an ensemble of 
twist-2 operators. These methods typically use either a large-momentum transfer to the currents that generates an operator product expansion (OPE) or short distance factorization (SDF) directly in coordinate space. These methods include: 
\begin{itemize}
    \item {\it Hadronic tensor approach.} The physical hadronic tensor $W^{\mu\nu}(q,\nu)$ ($q$ is the four-momentum flowing through currents, $\nu=P\cdot q$) can be reconstructed by analytical continuation in $\nu$ or time through Bayesian or maximum entropy {\it etc.} methods~\cite{Liu:1993cv,Liu:1999ak}. In the Bjorken limit, in which the momentum flowing through the currents asymptotically approaches the light-cone, the structure functions $W_i$ scale to the PDFs { such as $f_q(x)$}. The main challenge in this approach is the inverse spectral decomposition problem from the Laplace transformation in recovering large $\nu$.  
    It will require small lattice spacing~\cite{Liang:2019frk,Liang:2020sqi}.
    
\item {\it Compton amplitude approach or ``OPE without OPE''}. The physical Compton tensor $T^{\alpha\beta}(q,\nu)$ can be calculated through analytical continuation in the region where $\omega=2\nu/Q^2\le 1$~\cite{Aglietti:1998ur,Ji:2001wha}. Through a fixed-$Q^2$ dispersion relation, one can relate this amplitude to an integral over $W^{\alpha\beta}(\nu)$. A Taylor expansion of $T^{\alpha\beta}$ in the complex plane of $\omega$ around the origin (OPE) generates all the moments of physical structure functions or PDFs in the large-$Q^2$ limit, and the radius of convergence is at $|\omega|=1$. Therefore, one can, in principle, invert the dispersion integral to extract the PDFs~\cite{Chambers:2017dov} if the Compton tensor at large $Q^2$ is known for all $\omega\in [0,1]$, which is a challenge for $\omega \sim 1$.   A variation of the above approach is to use intermediate scalar or heavy quark to improve the convergence of the OPE to get the lowest few moments~\cite{Aglietti:1998ur,Detmold:2005gg,Detmold:2021uru}. 
    
\item {\it Current-current correlators}. Current-current correlators in Euclidean space provide a short distance factorization in the tower of twist-2 operators when $z^2\ll 1/ \Lambda_{\rm QCD}^2$~\cite{Braun:2007wv,Ma:2017pxb}. The coordinate space correlation in $\lambda = P\cdot z$ at fixed small $z^2$ 
can be matched to a sum of terms proportional to convolution of perturbatively calculated kernels with
physical PDFs. This approach to PDFs avoids engineering a large momentum transfer through currents, as required for the Compton or hadron tensor approaches. The method can also be naturally used to calculate higher moments of PDFs or amplitudes through Taylor expansion at small $\lambda$~\cite{Bali:2018spj,Joo:2020spy,Gao:2020ito}. To obtain the $x$-space PDFs from a limited range of $\lambda$, which arises because $z$ is restricted to perturbative distances at a finite $P$, requires solving an inverse problem like extracting PDFs from experimental data with limited collision energies and range of momentum transfer~\cite{Sufian:2020vzb}. 

\item{\it Pseudo-PDF approach}. SDF can also be used for Wilson-line operators~\cite{Ma:2014jla}  which are made of two spatially-separated QCD fields connected by straight Wilson lines in Eq. (4)~\cite{Ji:2013dva}, and the $x$-space matrix element with fixed $z^2$ has been called the pseudo-PDF~\cite{Radyushkin:2017cyf}. Features and limitations using Pseudo-PDFs are similar to the current-current correlators above, except might be computationally cheaper and have smaller finite volume effects.
\end{itemize}
All of these approaches provide constraints on collinear PDFs beyond individual moments. The lattice data on a range of coordinate-space correlations (``Ioffe-time distribution'') can be used to constrain PDFs through solving the inverse problem: either through a parametrization or neural network fit as in the standard phenomenological extraction of PDFs from hard scattering data. They can also be used together with experimental data to provide start-of-art constraints on partonic observables (see Sec.~VI). 

\subsection{Large-Momentum Expansion for Parton Observables}

An alternative approach to parton physics on the lattice follows from Feynman's original conception of partons as constituents of hadrons travelling at the speed of light~\cite{Bjorken:1969ja}. In this
formulation, parton observables are the physical 
observables of quarks and gluons in hadrons with  $P_\infty = P_z\to \infty$. For example, the collinear quark PDFs can be
written as~\cite{Ji:2013dva} 
    \begin{equation}
     f_q (x) \sim  \int d^2\vec{k}_\perp n_q (k^z=xP_\infty, \vec{k}_\perp)=\frac{1}{2}
     \int \frac{d\lambda}{2\pi}
     e^{ix\lambda} \langle P_\infty |\overline{\psi}_q(0) 
     \gamma^0W(0,z) \psi_q(z)|P_\infty\rangle \ , 
\end{equation}
where $n_q(\vec{k})$ is the ordinary time-independent momentum distribution function of a quark in the hadron and can be directly calculated on a Euclidean lattice. The Fourier transformation of the spatial correlation distance $\lambda = \lim_{P_z\to \infty, z\to 0} zP_z$ yields the momentum fractions $x$. This 
approach is entirely equivalent to the standard twist-2 operator definition above ~\cite{Izubuchi:2018srq}, but
applies to all parton observables including the collinear PDFs, TMD-PDFs and light-front wave functions, which are Fourier transformations of pure spatial (Euclidean) correlations $
    \langle P_\infty|O(\vec{z}_1,...,\vec{z}_n)|P_\infty\rangle$ and  $\langle 0|O(\vec{z}_1,...,\vec{z}_n)|P_\infty\rangle,$
where $\vec{z}_i$ are spatial coordinates of fields. The infinite-momentum limit is taken before ultraviolet (UV) regularization is imposed, and as such, partons in field theories with UV divergences are effective objects in the sense that they require an additional UV renormalization procedure to define. Usually, dimensional regularization and (modified) minimal subtraction is used to define the UV properties of partons.  

In practical calculations, a finite large momentum $P_z$ is used to approximate $P_\infty$,
and a large-momentum expansion is carried out, with systematic power corrections characterized by the expansion parameter, $\Lambda_{\rm QCD}^2/(x P_z)^2$. This follows from the physical picture of partons, which loose their meaning if their longitudinal momentum reaches the soft 
non-perturbative scale 
$\Lambda_{\rm QCD}$ 
(zero or soft modes instead of collinear ones). More importantly, the UV cut-off $\Lambda_{\rm UV}$ is always taken to 
be much larger than $P_z$, and the lattice matrix elements have to be matched to the standard light-cone parton distributions to account for different UV behavior. This approach to partons is similar to heavy-quark effective theory in which heavy-quark masses are taking to infinity, and has been called large momentum effective theory or LaMET~\cite{Ji:2014gla,Ji:2020ect}. 

\section{Theory and Computational Challenges}
\label{sec:TheoryChallenges}

Computing lattice matrix elements suitable for parton physics faces a number of challenges beyond the broad requirements of most lattice calculations for small lattice spacing, large volume, and physical pion mass. First and foremost is the large momentum necessary in hadron/Compton tensor currents $(q_\mu)$ or in hadron states ($P_\mu$). The second, relevant to Wilson-line operators, is renormalization of Wilson-line operators that have linear divergences requiring high-precision control of UV physics. The third is the long-range correlations in coordinate space necessary for small and large $x$ partons and for TMD impact-parameter space calculations in LaMET.  Fourth, in extracting $x$-dependent PDFs in SDF approaches, reliably estimating systematics in inverse problem is a challenge.  Finally, in calculations involving large $P_z$ and $z$, gluonic observables, and quantities depending on multi-variables such as GPDs and TMDs, poor signal-to-noise ratios ensure that significant computational resources are required, and thus new methods that can accelerate the standard calculations need to be systematically developed.  
Before getting into more details, we comment that most lattice calculations of PDFs in both SDF and LaMET used NLO matching or equivalently the NLO Wilson coefficients~\cite{Xiong:2013bka,Ma:2014jla,Ji:2017rah, Ji:2020ect}. However, also
NNLO matching exists~\cite{Chen:2020ody,Li:2020xml}, and has been used in
the lattice calculations of the valence pion PDF~\cite{Gao:2021dbh}. Two-loop matching are not yet widely available. 

\subsection{Large momentum currents and states}

Large momentum is essential for parton calculations on lattice in all new approaches (in contrast to moment calculations). The necessary condition for large momentum a fine lattice spacing. The smallest lattice spacing used in the current calculations are on the order of 0.04 fm~\cite{Izubuchi:2019lyk,Fan:2020nzz,Gao:2020ito}.  
The large-momentum directly controls the expansion parameter in LaMET approach and the range of calculable-$x$, and enables spatial correlations in SDF to extend to large $\lambda$. However, even on a fine lattice, large-momentum states can still pose a number of challenges: First of all,
the excited state contamination becomes severe as the energy difference between neighbouring states diminishes at large momentum. Moreover, the signal-to-noise ratio deteriorates at large momentum, due to statistical fluctuations.

Momentum smearing has been widely used to create large-momentum hadron states up to 3~GeV~\cite{Bali:2016lva}, and with sufficient statistics it may be possible to reach 5~GeV for nucleons. A large-momentum method has also been suggested for dilute stochastic sources~\cite{Wu:2018tvt}. To reach very small or large-$x$ partons, new methods are perhaps needed to create even larger momentum states. For large momentum states, the discretization errors could be enlarged to $a \gamma$ due to boost effects, which can most easily be seen in the momentum-energy dispersion relation. 

\subsection{Renormalization of linear divergences} 

Wilson-line operators contain linear divergences that must be subtracted before the continuum limit can be taken.  
Given the multiplicative renormalizability of the Wilson-line operator~\cite{Ji:2017oey,Green:2017xeu,Ishikawa:2017faj}, several schemes have been proposed to renormalize the corresponding quasi-PDF matrix
element. They can be divided into three types. The first type is to determine the Wilson line mass renormalization and the endpoint renormalization factors separately~\cite{Ishikawa:2016znu,Chen:2016fxx,LatticePartonCollaborationLPC:2021xdx}. The second type is to divide by the matrix element of the same operator in a different state, which can be chosen as the off-shell parton state in an RI-MOM scheme~\cite{Constantinou:2017sej,Alexandrou:2017huk,Chen:2017mzz,Stewart:2017tvs}, the zero-momentum hadron state in the ratio scheme~\cite{Orginos:2017kos}, the boosted hadron state in the ratio scheme~\cite{Fan:2020nzz}, or the vacuum state in the VEV scheme~\cite{Braun:2018brg,Li:2020xml}. The third type is a combination of the first two, such as the hybrid scheme~\cite{Ji:2020brr}.

RI-MOM schemes introduce additional operator mixing, because of the off-shell external states. This makes the situation particularly complicated when gluon PDFs are considered, and is possibly related to the recent observation that it fails to fully remove the linear divergence of the bare matrix element~\cite{Zhang:2020rsx}. In addition, it shares the same drawback as the ratio or the VEV scheme that the matrix element used in the renormalization
procedure is non-perturbative at large $z$, which invalidates the perturbative matching. The hybrid scheme
overcomes this problem~\cite{Ji:2020brr}, which is
equivalent to the ratio scheme for
$z<z_S$ (with $z_S$ being in the perturbative region, say smaller than $\sim0.3$~fm),
while for $z>z_S$ it is equivalent to the Wilson line mass renormalization scheme. The relevant UV divergence, in particular, the linear divergence, can either be calculated using the self renormalization procedure~\cite{LatticePartonCollaborationLPC:2021xdx} 
or from the calculation
of the static quark potential and the static quark free energy~\cite{Zhang:2017bzy,Gao:2021dbh}.

The linear divergence contains a renormalon uncertainty that is cancelled by $z$-dependent matching coefficients. Thus, any mass-subtraction procedure must be accompanied by properly renormalon-regularized matching coefficients.

\subsection{Methods for signal-to-noise improvement}

One of the biggest challenges for lattice calculations of correlation functions is the statistical noise at large Euclidean times, and, for non-local operators, at large space-like field separations. These challenges are exacerbated for baryons, which have significantly poorer signal-to-noise ratios than mesons, and at large external hadron momentum $P_z$. Extracting the best signal-to-noise ratio from the computationally expensive gauge configurations requires sampling algorithms that sample the configurations to the greatest degree possible. A widely adopted approach is that of low-mode averaging~\cite{Morgan:2004zh,DeGrand:2004wh}, in which the low modes of the Dirac operator are included exactly, whilst the higher modes are computed stochastically. Many hadronic quantities are not dominated by the low-lying eigenvalues of the Dirac operator and for these cases, all-mode averaging~\cite{Blum:2012uh,Shintani:2014vja} is a more efficient approach. The idea is to use covariant symmetries to construct approximate correlation functions, which are cheaper to compute, to calculate expectation values without introducing any systematic bias.

Smearing, such as the widely employed Jacobi~\cite{UKQCD:1993gym} or Wuppertal~\cite{Gusken:1989qx} methods, have proved an invaluable tool to improve the signal-to-noise ratio for interpolating operators, and the momentum-smearing method~\cite{Bali:2016lva} has been particularly important for hadrons with non-zero momentum. The distillation framework~\cite{HadronSpectrum:2009krc} is a further example of smearing with the important benefit that it enables a momentum projection to be performed at each point in a two-point or three-point function, thereby providing a more complete sampling of the lattice than is possible with the Jacobi or Wuppertal frameworks. The efficacy of combining the more complete sampling of the lattice made possible by distillation with the ability to reach higher momenta using momentum smearing ~\cite{Egerer:2020hnc} has been demonstrated in the recent calculation of the unpolarized PDF using distillation~\cite{Egerer:2021ymv}, which showed improved precision in comparison with an earlier work employing Jacobi smearing~\cite{Joo:2019jct}.

{Another way to reduce the noise is to consider smaller separations between the hadron source and the hadron sink. For small source-sink separation the excited state contamination is significant, but using 
several source-sink separations, different types of sources it is possible to parametrize the contribution
of excited states in a reliable manner~\cite{Izubuchi:2019lyk,Gao:2020ito,Egerer:2021dgg}. 
Adding more source-sink separations in the analysis increases the cost of computation, but this increase
is linear compared to the exponential increase of the cost that is needed to improve the signal at large-source sink separations. In the future it may be interesting to explore the use of even more source-sink separation in such analyses.}

Gluon operators, in particular, suffer from significant signal-to-noise issues, and require gauge-field smearing techniques to improve the signal. These techniques include the gradient flow~\cite{Luscher:2010iy,Shanahan:2018pib,HadStruc:2021wmh}, stout smearing~\cite{Morningstar:2003gk,Alexandrou:2017oeh}, and HYP smearing~\cite{Hasenfratz:2001hp,Fan:2018dxu,Fan:2020cpa,Fan:2021bcr,Salas-Chavira:2021wui}. Further improvement can be obtained through the cluster-decomposition error reduction method~\cite{Liu:2017man}. For operators with signal that decays exponentially with spacetime distance, integrating over all spacetime points necessarily includes a region of spacetime that is predominantly noise and can be omitted from the integration region. CDER has been applied to the nonperturbative determination of the renormalization parameter relevant to the gluon momentum fraction~\cite{Yang:2018bft}.

\subsection{Method-Dependent Challenges}

\subsubsection{Inverse problems PDF extraction in SDF}

The determination of light-cone PDFs from lattice simulations is subject in general to an inverse problem. This is because the fields separation of the associated PDF correlation function is calculable in LQCD only away from the light-cone direction. Fortunately, off-the-light-cone correlation functions can be factorized as a convolution of light-cone PDFs and perturbative matching coefficients similar to factorization in hard processes in high-energy reactions. Since the perturbative matching coefficients are distributions defined only under integration, it is not possible to find a unique inverse function that deconvolutes the lattice correlation function. Moreover, the sensitivity on different momentum fractions requires to have sufficient resolution in the computation of the correlation function as a function of the field separation, which in practice is limited by the discrete nature of the lattice spacing and the lattice volume.  

In order to solve the inverse problem, it is required to make assumptions for the PDFs in regions of momentum fractions that are not sensitive to the lattice data. In this context, the Bayesian approach becomes the ideal framework that allows us to define suitable priors to solve the inverse problem. The explicit specification of the priors used becomes an important step in the solution of the problem. Here we recall some of the  Bayesian approaches found in the  literature.

\begin{itemize}
    \item {\it Backus-Gilbert method}. Given a discrete set of data points, one can use the Backus-Gilbert method to obtain the $x$-dependent PDFs by regularizing the singular inversion matrix~\cite{Karpie:2019eiq}. The result is equivalent to some mathematical extrapolation and interpolation of the discrete data points.  The dependence of the results on specific pre-conditions needs be quantified.  This method has been applied to the spectral decomposition of the hadronic tensor~\cite{Liang:2019frk}. This approach is found to better delineate broad structures. 

    \item {\it Maximum entropy method (MEM)}. The MEM method solves the inverse problem by adding reasonable physical constraints~\cite{Karpie:2019eiq}.
    This method and its improved version -- Bayesian reconstruction, are good at identifying discrete states while solving the inverse problem for the hadronic tensor~\cite{Liang:2019frk}.

    \item {\it Parametrization based analysis}. In this approach, one starts by modeling the PDFs using a flexible parametrization. These models range from simple traditional parametrizations inspired by Regge-physics and spectator counting rules to more flexible parametrizations that can be realized via neural-net approach \cite{Karpie:2019eiq,Cichy:2019ebf,DelDebbio:2020rgv,Zhang:2020gaj,DelDebbio:2021whr}. Regardless of the modeling, the Bayesian posterior provides a mechanism to implement uncertainty quantification for the PDFs. This approach offers the possibility to combine lattice and experimental information within a unified Bayesian inference framework.~\cite{Bringewatt:2020ixn} 
\end{itemize}

\subsubsection{Reaching long-range correlations in LaMET}

LaMET performs calculations of the PDFs at a chosen $x$ from the momentum densities of quarks and gluons in a large-momentum hadron. This approach
does not have the inverse problem described in the 
previous subsection, but requires information for spatial correlations at all $z$-distance in order to carry out the Fourier transformation to $x$-space. Because of QCD color confinement (or the absence of massless excitations) in lattice simulations, all spatial correlation functions should decay exponentially at large distances, with the smallest decay exponent corresponding to the lowest hadron mass of appropriate quantum numbers. In the case of non-singlet quark PDFs, this is a light-quark and static-color-source bound state. After subtracting the linear-divergent mass counter-term with an IR renormalon regularization, the decaying exponential corresponds to the light-quark binding energy of order $\Lambda_{\rm QCD}$.
Therefore, lattice correlation functions must have good signal up to the exponential decay region between 1 and 1.5~fm before a reliable LaMET analysis can be carried out. For very small-$x$ partons, new methods for calculating longer range correlations must be developed.

\section{Benchmarking and Precision Calculations}
\label{sec:PrecisionPDF}
 
In kinematic regions where experimental data are plentiful or overconstrained, such as the mid-$x$ region of the PDFs, there is consistency among different PDF sets. Thus any successful lattice calculations must be calibrated against the precision phenomenological PDF sets, as well as the well-established moment approach.

Extensive calculations have been carried out for isovector PDFs and DAs using the Wilson-line operator, because they are the easiest to implement. The twist-2 spin-independent, helicity, and transversity PDFs of the nucleon and the pion have been calculated in both LaMET and SDF. Calculations have been performed using several lattice spacings, which allows results to be extrapolated to the continuum, albeit at an unphysical pion 
mass~\cite{Izubuchi:2019lyk,Gao:2020ito,Alexandrou:2020qtt,Lin:2020fsj,Zhang:2020gaj}. 
Development of 5\% level or better precision for isovector collinear PDFs for $0.2 < x < 0.6$ and with improved determination sea quark distributions and large-$x$ quark and gluon distributions, may have important impacts in high-energy collider experiments.  

For collinear PDFs, the proton PDFs from global analyses can serve to calibrate lattice calculations. With systematic uncertainties, such as excited-state contamination, renormalization and continuum limit, physical pion mass, finite volume effects, perturbative matching and power corrections under control, one should expect agreement between lattice determinations and global analyses in the kinematic regions that are robustly constrained by the data. The flavor-singlet quark PDF, however, are more challenging as one has to include the disconnected diagrams and mixing with the gluon PDFs, so a more reasonable target precision in this case may be 10-20\% in practice.

\begin{itemize}
    \item {\it Excited state contamination}
The need to separate the matrix elements of the ground-state hadron from the contributions of excited states that inevitably contribute to the computed correlation function is one of the most severe challenges in precision lattice calculations.  The challenges become increasingly severe as the spatial momentum of the states is increased.  First, for a given spectrum of states, the relative separation between their energies narrows with increasing momentum.  Second, the symmetry group governing the spectrum reduces to a smaller little group, thereby increasing the number of states contributing to any correlation function. There have been numerous efforts both to control the contribution of excited states to ground-state matrix elements~\cite{Yoon:2016dij,Park:2021ypf,Lin:2020rxa}. Two approaches have been adopted to address these issues.  The first aims to attenuate the time dependence of the contribution of excited states through the use of the summation method.  The second approach aims to reduce the leading energy dependence of the excited-state contributions through the use of the variational method and the solution of a generalized eigenvalue problem. The amalgam of these two approaches then leads to the summed Generalized-Eigenvalue (sGEV) approach ~\cite{Bulava:2011yz}.  The goal is to enable the reliable extraction of hadronic matrix elements with controlled contributions from excited states from correlation functions at sufficiently short temporal separations that the signal-to-noise ratio remains acceptable.

The application of the variational method requires a basis of operators be constructed that respects the symmetries of the lattice, including for states in motion~\cite{Thomas:2011rh},  and here the use of the related distillation~\cite{HadronSpectrum:2009krc} or stochastic LaPH~\cite{Morningstar:2011ka} frameworks affords a computationally efficient means of implementing such as basis.  Notably, both frameworks admit the straightforward inclusion of multi-hadron operators into the basis whose contribution is expected to be notably significant in the case of the nucleon form factors~\cite{Bar:2019igf,Bar:2021crj}. This framework together with the use of the variational method has been applied to the calculation of the nucleon charges both for hadrons at rest~\cite{Egerer:2018xgu}, and a non-zero spatial momentum~\cite{Egerer:2020hnc}.  The need for such approaches is particularly crucial for quantities such as the flavor-singlet and gluonic distributions where the signal-to-noise ratio is already poor, and key to the high precision obtained in the calculation of the gluonic matrix elements of Ref.~\cite{HadStruc:2021wmh}.

\item {\it Renormalization and continuum extrapolation} 
There have been some studies of the continuum extrapolation of the quasi-PDF method in the pion and kaon distribution amplitudes~\cite{Zhang:2020gaj} and in nucleon PDFs~\cite{Alexandrou:2020qtt};
both cases use three lattice spacings but a single heavy quark mass with $M_\pi > 300$~MeV.
Ref.~\cite{Lin:2020ssv} determines valence-quark PDFs of the pion and kaon using two lattice spacings (0.06 and 0.12~fm) and 3 pion masses ($M_\pi \in [220,690]$~MeV).
This work is the first study of lattice PDFs to take the continuum-physical limit of the matrix elements with a sufficient number of lattice spacings and light pion masses, an important step toward precision PDFs from lattice QCD.
Ref.~\cite{Lin:2020fsj} calculated isovector nucleon PDF using three lattice spacings $a \in [0.06,0.12]$~fm, pion mass $M_\pi \in [135,318]$~MeV and box size $L\in [2.9,5.5]$~fm (which make $M_\pi L\in [3.3,5.5]$). Both $O(a)$ and $O(a^2)$ continuum extrapolation are performed in the calculation with nucleon boost momentum around $2.2$ and 2.6~GeV; the final continuum-limit results are combined according to Akaike information criterion (AIC), with reasonable  agreements for mid- to large-x regions, and compatible within 2 standard deviations for $x < 0.4$. The nucleon isovector moments $<x^n>$ are around 0.2, 0.06, and 0.04 for $n = 1, 2, 3$, respectively. Work are still need to reduce the twist-4 systematics and improve the lattice determination of small-$x$ and antiquark PDFs. 
Another continuum limit calculation has also been done in the SDF approach~\cite{Karpie:2021pap}, where three lattice spacings, 0.048, 0.065, and 0.075~fm have been used in a two-flavor QCD calculation, all with pion mass around 440~MeV. Using the ratio approach to renormalize the correlation functions at small $z$, the lattice-spacing dependence can be controlled very nicely. No noticeable variation in $a$ has been found in the calculation.

There has also been a first next-to-next-to-leading order (NNLO) accuracy~\cite{Gao:2021dbh} lattice calculation of the pion valence quark PDF with a reliable determination of the PDF for $0.03\lesssim x \lesssim 0.80$ with 5-20\% (mostly statistical) uncertainty at pion momentum $2.42$ GeV
In this calculation, the lattice results at different pion momenta start to converge at moderate $x$ when $P_z\ge 1.45$ GeV, which corresponds to a Lorentz boost factor of $\sim 5$.
Therefore, for the proton PDF one may expect a 5\% lattice determination for $0.1\lesssim x \lesssim 0.8$ with proton momentum $\sim 5$ GeV at physical pion mass. Since the isovector PDF is the most studied quantity so far, it is likely to reach this target precision within the next few years. The agreement with global analysis in the antiquark region is particularly important as at $x=0.1$ the flavor-asymmetry of the antiquark distribution already becomes important.

Continuum extrapolation is also important for LaMET due to potential operator mixing in the nonlocal operators. The nonlocal operators for the quasi-PDFs can mix with a tower of higher-dimensional operators at $\mathcal{O}(a)$, even if all symmetries (including chiral symmetry) are restored~\cite{Green:2017xeu,Chen:2017mie,Green:2020xco}.  This is different from the situation for local operators, where mixing can occur at $\mathcal{O}(a^2)$ if a chiral lattice fermion action is used. To ensure that such operator mixings do not contaminate the final results of the lattice PDF calculations, it is important to take the continuum limit. 

\item {\it Pion mass}. 
To make direct comparison with experimental measurements or globally fitted PDFs, lattice calculations must either be directly calculated at or reliably extrapolated to the physical pion mass.
Since the computational cost of evaluating the fermion contribution to the path integral increases with a large inverse power of the quark mass, many lattice-QCD calculations are performed at unphysically heavy pion masses.
In recent years, with the help of hardware and software advancements, there are increasingly many results published using direct calculation at the physical pion masses.
However, such direct calculations still yield larger statistical uncertainty. 
For example, Ref.~\cite{Joo:2020spy} studies nucleon isovector matrix elements using SPDF method with $M_\pi \in[172,358]$~MeV;
the result at their lightest pion mass has much larger errors compared with heavier pion mass (Fig.~1 in  Ref.~\cite{Joo:2020spy}).
No pion-mass dependence is observed outside the uncertainty in this study. 
The isovector nucleon matrix elements calculated also shows mild pion mass dependence, as seen in Fig.~4 in Ref.~\cite{Lin:2020fsj} with $M_\pi \in [130,310]$~MeV at small Wilson-line displacement with LaMET method. 
Similar behavior is also observed in the valence-quark PDFs of the pion and kaon in Ref.~\cite{Lin:2020ssv} 
for pion mass below 310~MeV, and that of pion from  
LSC method~\cite{Sufian:2020vzb}.  
There is no surprise here;
even with the traditional local-operator moment method, having accuracy of 10\% error or larger, the calculation at different pion masses usually agrees within 1--$2\sigma$ (see, for example, Figure~B.1 in Ref.~\cite{Lin:2017snn}).
However, tension occurs when the precision is improved to the few-percent level, such as in the case of axial charge (as seen in Fig.~3 of Ref.~\cite{Chang:2018uxx}).
Future study with improved precision in $x$-dependent methods will shed more light on the role of pion mass in PDF calculations. 

\item {\it  Finite volume}. To further improve the lattice computations at physical pion mass, the remaining lattice systematics must be treated, by extrapolation to infinite volume and the continuum limit. The possibility of enhanced finite volume effects for non-local operators was first studied in~\cite{Briceno:2018lfj}, but numerical evidence suggests that the finite-volume effects for Wilson-line operators are negligible at current lattice precision. In the quasi-PDF approach,
a first study of finite-volume systematics was undertaken in Ref.~\cite{Lin:2019ocg} with isovector both polarized and unpolarized nucleon PDFs;
three lattice volumes (2.88, 3.84, 4.8~fm) were studied at pion mass 220-MeV and nucleon momenta 1.3 and 2.6~GeV, and no noticeable finite-volume dependence was found. This is consistent with a later study in chiral perturbation theory (ChPT)~\cite{Liu:2020krc}, which showed that momentum boost reduces the finite-volume effect, since the length contraction of the hadron makes the lattice effectively bigger. ChPT also showed that for nucleon momenta greater than 1~GeV and the lattice size times pion mass greater than 3, then the finite-volume effect on the isovector nucleon PDF is less than $1\%$.
This conclusion is consistent with the numerical findings of Ref.~\cite{Lin:2019ocg} and with the study of finite volume effects in the Wilson-line renormalization parameter in Ref.~\cite{Alexandrou:2019lfo}. Some evidence for non-negligible finite volume effects in current-current correlators was found in Ref.~\cite{Sufian:2020vzb}.

\item {\it Perturbative Matching}. 
In most existing lattice calculations, the perturbative matching has been implemented up to NLO, and the missing higher-order contribution is an important source of systematic uncertainty. So far, the NNLO matching is only available for the unpolarized isovector quark PDF~\cite{Chen:2020ody,Li:2020xml}, both in coordinate space and in momentum space. The calculation is done in the RI/MOM and the VEV scheme, thus the result needs to be adapted so that it can be applied to hybrid or self-renormalized matrix elements. In Ref.~\cite{Gao:2021dbh}, the pion valence quark distribution was extracted up to NNLO in the hybrid scheme, where the NNLO matching effect can reach $\sim 10\%$ in the moderate $x$ region. There is also evidence that the NNLO effect is of similar size for the meson DA~\cite{Hua:2022kcm}. To target a $5\%$ precision, it is desirable to go beyond NNLO. In Ref.~\cite{Braun:2020ymy}, the anomalous dimension of the spatial quark bilinear operator has been calculated to three-loop accuracy, which is independent of the Dirac structure in the operator. The calculation is also done for the gluon operator to two-loop accuracy in the same paper. Besides, a subset of diagrams with fermion bubble chains has been worked out to all-orders in perturbation theory in Ref.~\cite{Braun:2018brg}. The techniques used in these calculations can be helpful to the calculation of the NNLO matching. Moreover, the use of hybrid or self-renormalization also facilitates the calculation, as one needs to work with dimensional regularization and on-shell states only. We expect the NNLO matching for all leading-twist partonic quantities, including the PDFs, GPDs, DAs, and even TMDs and light-front wave functions to be available in the next few years. 

\item {\it Calibrating with Mellin moments}. The first important check for $x$-dependent lattice PDF calculations at the continuum limit is to reproduce moment results. There has been an attempt to rate lattice calculations beyond the nucleon charges collected in the FLAG report~\cite{Aoki:2021kgd}, and recent surveys of the lattice moments and PDF calculations and their benchmarks in the nucleon sector can be found in Ref.~\cite{Lin:2017snn,Constantinou:2020hdm}. The coordinate-space correlations can be Taylor expanded near the origin; the coefficients are the moments of the PDFs. One can use the direct calculations of the moments to calibrate the correlation
functions near the origin. 

\item {\it Computational Challenges}. 
Summarising the various sources of systematic errors that have been
discussed in the previous subsections, we can estimate the size 
of simulations that would allow a comparison with global analyses at 
the 5\% level.  
In order to perform a controlled extrapolation to the continuum limit
at least three lattice spacings are desirable; the discussion above on 
renormalization and the continuum limit suggests that the finest lattice
spacing should be approximately $a=0.05~\mathrm{fm}$ (corresponding to 
 $a^{-1} \approx 4~\mathrm{GeV}$.) 
Even though the dependence on the pion mass seems to be mild, simulations
for precision benchmarks should aim to be run at the physical pion mass, 
$M_\pi=139~\mathrm{MeV}$. This in turn sets the size of the physical 
volume, a conservative estimate based on existing studies would suggest 
to achieve at least $M_\pi L = 3$, for which we would require lattices 
of linear size $L=4.5~\mathrm{fm}$ and therefore a number of lattice sites
$L/a\approx 90$. A safer option of $M_\pi L =4$ would increase the cost 
proportionally.  
To have a reliable calculation, we need nucleon momenta of $P\approx  2.6~\mathrm{GeV}$, as
discussed in earlier subsection, 
corresponding to $P a \simeq 0.65$. While there is some hierarchy between the momentum and the cutoff scale, this number may turn out to be too small to control lattice artefacts due to the 
nucleon momentum. 
Finer lattice spacings would involve new challenges, related to the 
critical slowing down of topological modes, and would result in lattice 
with an inversely-proportional larger number of sites. 
These type of
simulations will be perfect candidates for exascale computations on 
forthcoming machines.

\end{itemize}

If the agreement is met for both unpolarized and helicity isovector PDFs with 5\% precision, then we will have much confidence to use the lattice predictions for the transversity PDFs, as well as all PDFs in $x$ regions or parton flavors where experiments have not constrained so well. Besides, since the method for calculating GPDs is analogous to that for the PDFs except for the dependence on extra kinematic variables, we can also expect the lattice predictions to be reliable, which will offer much needed nonperturbative inputs for the experiments at COMPASS II, JLab 12 GeV Upgrade and EIC.

\section{New Partonic Observables}
\label{sec:femtography}

Lattice parton methods have been extended beyond the standard twist-2 collinear parton distributions. In particular, they can be used to calculate GPDs and TMDs, as well as higher-twist collinear PDFs. 

\subsection{Generalized Parton Distributions (GPDs)}

Generalized parton distributions provide hybrid momentum and coordinate space distributions of partons and bridge the standard nucleon structure observables: form factors and collinear PDFs.
More importantly, GPDs provide information on the spin and mass structure of the nucleon.  
GPDs bring the energy-momentum tensor matrix elements within experimental grasp through electromagnetic scattering and can be viewed as a hybrid of parton distributions (PDFs), form factors, and distribution amplitudes.
For example, the forward limit of the unpolarized and helicity GPDs lead to the $f_1(x)$ and $g_1(x)$ PDFs, respectively.
Taking the integral over $x$ at finite values of the momentum transfer results in the form factors and generalized form factors.
In the case of the unpolarized GPDs, for example, one obtains the Dirac ($F_1$) and Pauli ($F_2$) form factors.
Several of these limits of the GPDs have physical interpretations, for instance, the spin decomposition of the proton using Ji's sum rule~\cite{Ji:1996ek}.

Information on GPDs from lattice QCD has been available via their form factors and generalized form factors, using the operator product expansion (OPE).
As in PDFs, such information is limited due to the suppression of the signal as the order of the Mellin moments increases and the momentum transfer between the initial and final state increases.
Significant progress has been made towards new methods to access the $x$- and $t$-dependence of GPDs ($t=-Q^2$), which is driven by the advances in PDF calculations. 
In lattice QCD, there are several challenges in calculating GPD using these new methods.
The extraction of GPDs is more challenging than collinear PDFs, because GPDs require momentum transfer, $Q^2$, between the initial (source) and final (sink) states.
Another complication is that GPDs are defined in the Breit frame, in which the momentum transfer is equally distributed to the initial and final states;
such a setup increases the computational cost, as separate calculations are necessary for each value of the momentum transfer.

The first lattice $x$-dependent GPD calculations were carried out in Ref.~\cite{Chen:2019lcm}, studying the pion valence-quark GPD at zero skewness with multiple transfer momenta 
with pion mass $M_\pi \approx 310$~MeV.
There is a reasonable agreement with traditional local-current form-factor calculations at similar pion mass, but the current uncertainties remain too large to show a clear preference among different model assumptions about the kinematic dependence of the GPD.
There has also been recent progress made in lattice QCD to provide the Bjorken-$x$ dependence of the isovector nucleon GPDs, $H$, $E$ and $\tilde{H}$. 
Ref.~\cite{Alexandrou:2020zbe} used LaMET to calculate both unpolarized and polarized nucleon isovector GPDs with largest boost momentum 1.67~GeV at pion mass $M_\pi \approx 260$~MeV with one momentum transfer.
This work also presented results at nonzero skewness, with additional divergence near  $x=\xi$  due to the matching.
Refs.~\cite{Lin:2020rxa,Lin:2021brq} reported the first lattice-QCD calculations of the unpolarized and helicity nucleon GPDs with boost momentum around 2.0~GeV at physical pion mass with multiple transfer momenta, allowing study of the three-dimensional structure and impact-parameter--space distribution.
Results for the moments of the integral of the $H$, $E$ and $\tilde{H}$ GPDs extracted from the lattice are within a couple sigma of previous lattice calculations  using OPE operators from traditional form factors and generalized form factors at or near the physical pion mass.
Such lattice inputs can provide useful constraints 
to the best determination of physical quantities using both theoretical and experimental inputs. 

\subsection{Transverse Momentum Distributions (TMDs) }
\label{sec:tmd}

Transverse momentum distributions measure the parton transverse momentum $k_T$ with longitudinal momentum fraction $x$, and are nonperturbative inputs for processes that follow TMD factorization, such as Drell-Yan and semi-inclusive DIS (SIDIS). When $k_T\gg \Lambda_{\rm QCD}$, TMDs can be perturbatively matched onto the collinear PDFs, and thus are predictable from the latter. But when $k_T\sim \Lambda_{\rm QCD}$, the TMDs are intrinsically nonperturbative and must be determined from experimental measurement or first-principles calculations. TMDs depend not only on the renormalization scale, but also on the so-called Collins-Soper (CS) scale, which is related to the parton energy. The anomalous dimension for evolution in the CS scale, which is also known as the CS kernel, depends on $b_T$, the Fourier conjugate of $k_T$, and becomes nonperturbative when $b_T\sim 1/\Lambda_{\rm QCD}$.
At LHC, the Drell-Yan differential cross sections have been measured at sub-percent level precision for final-state transverse momentum $p_T$ as small as $1$ GeV~\cite{CMS:2019raw,ATLAS:2019zci}, which is sensitive to the nonperturbative TMDs. The lack of knowledge of the CS kernel and the TMD in the nonperturbative region limits theory predictions at low $p_T$. 
The JLab 12 GeV and EIC will also provide a significant amount of SIDIS data in the low-energy region sensitive to $k_T\sim \Lambda_{\rm QCD}$.
Therefore, a systematic lattice QCD determination of these quantities will provide opportunities for further precision tests of the Standard Model with LHC data and mapping the instrinsic transverse-momentum structure of the proton at JLab 12 GeV and the EIC.

Like PDFs, TMDs are defined in terms of light-cone correlation functions, and they are challenging to calculate directly in lattice QCD. However, the LaMET framework~\cite{Ji:2013dva,Ji:2014gla,Ji:2020ect} provides a promising pathway towards the determination of TMDs by matching calculations of equal-time correlation functions in large-momentum hadron states, or the quasi-TMDs, to the light-cone correlation functions defining TMDs~\cite{Ji:2014hxa,Ji:2018hvs,Ebert:2018gzl,Ebert:2019okf,Ebert:2019tvc,Ji:2019sxk,Ji:2019ewn,Vladimirov:2020ofp}. This, and related, approaches have been implemented in recent years to provide the first first-principles calculations of aspects of TMDs~\cite{Musch:2010ka,Musch:2011er,Engelhardt:2015xja,Yoon:2016dyh,Yoon:2017qzo}.

\paragraph{Collins-Soper kernel.}
Determinations of the CS kernel from experimental data via simultaneous global fits with TMDs reveal some discrepancies between different analyses, in particular in the region $q_T\le 500$ MeV~\cite{Bacchetta:2017gcc,Scimemi:2017etj,Bertone:2019nxa,Scimemi:2019cmh,Bacchetta:2019sam,Vladimirov:2020umg}. Improvements to the systematic understanding of the kernel via first-principles calculations have the potential to significantly improve this situation, and thereby enable more precise determinations of TMDs and related processes. Refs.~\cite{Ji:2014hxa,Ebert:2018gzl,Ebert:2019okf,Ebert:2019tvc,Vladimirov:2020ofp} recently demonstrated how the CS kernel can be determined in lattice QCD from calculations of the quasi-TMDs, and this approach, and variations thereof, has been implemented in Refs.~\cite{Shanahan:2020zxr,Zhang:2020dbb,Schlemmer:2021aij,Li:2021wvl,Shanahan:2021tst}.
 
With sufficient investment of computational resources, it is likely that the CS kernel can be determined nonperturbatively via lattice QCD calculations with few or sub-percent precision at $b_T\lesssim 0.5$~fm or $k_T\gtrsim 0.4$ GeV in the coming decade. This region, importantly, overlaps with the nonperturbative region that is relevant for constraining global fits of TMDs, and it can be expected that the precision of such future fully-controlled lattice QCD calculations will be sufficient to begin to differentiate the different behaviors observed in models of the kernel in this region and ultimately will be used as input to phenomenological analyses.

\paragraph{TMDs and soft function.}
The physical TMDs that enter the Drell-Yan and semi-inclusive DIS factorization formulae include the beam and soft functions. Depending on how the rapidity divergences are regulated, there are different schemes to define the beam and soft functions separately, and those that are amenable to lattice calculations use off-the-light-cone Wilson lines in the TMD correlators, i.e., the Collins scheme. The Collins beam function can be easily obtained from the equal-time TMD correlator, or quasi beam function, in a large-momentum hadron state. The soft function, however, faces more challenges as it involves Wilson lines along two different off-the-light-cone directions, which makes it impossible to obtain it from an equal-time correlator through a single Lorentz boost.

Recently, a breakthrough was made within the LaMET formalism of the lattice QCD calculation of the soft function~\cite{Ji:2019sxk}. This method has been applied in exploratory lattice calculations~\cite{Zhang:2020dbb,Li:2021wvl}, where the first results show its robustness in obtaining the soft function for $b_T\lesssim 0.5$ fm or $k_T\gtrsim0.4$ GeV. With the soft function available, one can obtain the $x$ and $b_T$ dependence of the TMDs with the calculation of the quasi beam function~\cite{Ji:2019ewn}. In this way, the lattice QCD calculations can be used to predict the Drell-Yan cross sections at low $p_T$ for the LHC. Apart from the unpolarized TMDs, the LaMET approach can also be used to calculate the spin-dependent TMDs, such as the Sivers function~\cite{Ji:2020jeb}, which could have a significant impact on experiments at the JLab 12 GeV upgrade, RHIC and the EIC.

\paragraph{TMD wave function.}
Under light-cone quantization, the full information about the parton structure of a hadron, including PDFs, TMDs, GPDs and Wigner distributions, can be obtained from the hadron's Fock-space TMD wave function. However, the drawback of this formalism is that one needs to obtain the wave functions for an infinite number of Fock-space bases, and in practice a truncation is inevitable and not always under systematic control.
Currently, there is no experimental data on TMD wave functions, so a lattice QCD determination of even the lowest Fock-state wave function can provide much needed information. 
With the LaMET approach, the method for calculating the TMD wave function is similar to that for the TMDs~\cite{Ji:2021znw}, which involves the computation of the same soft function. First efforts to employ this method are being carried out, and as a byproduct one can also extract the CS kernel from them~\cite{Zhang:2020dbb,Li:2021wvl}.

The lattice results of these quantities will be particularly applicable to measurements of the Drell-Yan and semi-inclusive DIS processes at low $p_T$ from $\sim1$--10 GeV, which are sensitive to the nonperturbative TMDs and their CS evolution. Experiments at the JLab 12 GeV Upgrade and the future EIC will provide an enormous amount of data in this kinematic regime over the next two decades.
Global fits of TMDs have not reached the same consistency and precision as those of the PDFs, so benchmarking lattice calculations relies more on the theory itself. Fortunately, quantities such as the CS evolution kernel and the soft function are perturbatively calculable at $b_T\ll 1/\Lambda_{\rm QCD}$, which offers accurate calibration for lattice calculations. According to recent calculations~\cite{Shanahan:2020zxr,Zhang:2020dbb,Schlemmer:2021aij,Li:2021wvl,Shanahan:2021tst}, it is feasible to target a 5\% precision for $b_T\sim 0.1$ fm with contemporary available computing resources, which would require a very fine lattice $a\lesssim 0.03$ fm to suppress the discretization effects. Like the PDF calculations, this precision also requires large hadron momentum to suppress the power corrections of ${\cal O}(1/(P_zb_T)^2)$. Agreement between lattice results and perturbation theory within the small $b_T$ region, will all lattice systematics under control, will provide confidence in the predictions at larger $b_T$, which can be used as inputs for experimental analysis.

\subsection{Higher-Twist Observables}
\label{sec:beyondTwist3}

The discussion so far has focused on the leading twist contributions, which capture the most important components of the partonic structure of hadrons. In addition, twist-2 distribution functions are the easiest to isolate from experimental data sets. Twist-3 contributions do not have a probabilistic interpretation, but contain important information on quark-gluon correlations~\cite{Balitsky:1987bk,Kanazawa:2015ajw}. Twist-3 contributions also appear in QCD factorization theorems for various observables and some are related to properties such as the parton orbital angular momentum~\cite{Ji:1996nm}. Twist-3 distributions therefore play an important role in characterizing the parton structure of hadrons in novel ways.

At the twist-3 level, the proton has three PDFs: the chiral-even $g_T(x)$, and chiral-odd $e(x)$ and $h_L(x)$~\cite{Jaffe:1991kp, Jaffe:1991ra}. Semi-classically, $g_T(x)$ is connected to the transverse force acting on the active quark in a transversely polarized nucleon, while $e(x)$ is related to the transverse force acting on transversely polarized quarks in an unpolarized nucleon~\cite{Burkardt:2008ps}. Experimentally, $g_T(x)$ PDF can be extracted through the DIS structure function $g_2^{\rm s.f}$ (or, equivalently $g_T^{\rm s.f}$)~\cite{Flay:2016wie, Armstrong:2018xgk}, while $e(x)$ can be accessed in an unpolarized Drell-Yan process at twist-4 order~\cite{Jaffe:1991ra}, or through a twist-3 single-spin asymmetry in semi-inclusive DIS~\cite{Levelt:1994np}, or di-hadron production in electron-proton collisions~\cite{Bacchetta:2003vn}. Limited data from semi-inclusive DIS have been obtained by the HERMES and CLAS collaborations~\cite{HERMES:1999ryv,HERMES:2001hbj,CLAS:2003qum,CLAS:2014dmz}, and from di-hadron production in electron-proton collisions by the CLAS collaboration~\cite{Courtoy:2014ixa}.  $h_L$ is relevant to a twist-3 double-spin asymmetry in Drell-Yan~\cite{Jaffe:1991kp, Jaffe:1991ra, Tangerman:1994bb, Koike:2008du}. Despite the importance of twist-3 PDFs, their study experimentally is in its infancy. This situation worsens for twist-3 GPDs, where experimental data are basically non-existent. Lattice QCD therefore provides the most promising source of information for understanding hadron structure beyond leading twist. 

Recently, dedicated calculations of the $x$ dependence of twist-3 PDFs and GPDs were performed in lattice QCD using LaMET~\cite{Bhattacharya:2020cen,Bhattacharya:2021moj,Bhattacharya:2021oyr}. Calculations of twist-3 quantities are much more complex than twist-2 for a number of reasons. The first challenge is the presence of mixing between 2-parton twist-3 contributions and quark-gluon-quark contributions (genuine twist-3). As a result, the matching formalism discussed in Sec.~\ref{sec:TheoryChallenges} becomes a $2\times2$ mixing matrix in the twist-3 case. Unlike the case of twist-2, at twist-3 the matching contains singular zero-mode contributions, which have implications in Burkhardt-Cottingham sum rules~\cite{Kodaira:1998jn,Efremov:2002qh,Pasquini:2018oyz}. First efforts on the matching for twist-3 PDFs can be found in Refs.~\cite{Bhattacharya:2020jfj,Bhattacharya:2020xlt,Braun:2021aon,Braun:2021gvv}. Computationally, extraction of quark-gluon-quark correlation are complicated due to the presence of the gluon field strength tensor in the operator and two Wilson lines. This increases the computational cost as these distributions depend on two momenta when Fourier transformed. In addition, gluon operators have significantly decreased signal-to-noise ratio compared to the quark distribution functions.

The first lattice calculation of a twist-3 PDF was of $g_T(x)$, which serves as a proof-of-principle computation. The work is performed at pion mass of about 260~MeV and momentum boost up to 1.67~GeV~\cite{Bhattacharya:2020cen}. A calculation of the $h_L(x)$ PDF using the same ensemble and momentum boost can be found in Ref.~\cite{Bhattacharya:2021moj}. In these calculations, it was confirmed numerically that $g_T(x)$ and $h_L(x)$ are as sizeable as the $g_1(x)$ and $h_1(x)$ PDFs. One of the most interesting investigations regarding twist-3 PDFs is the Wandzura-Wilczek (WW) approximation~\cite{Wandzura:1977qf,Jaffe:1991ra,Jaffe:1991kp}, where the PDFs $g_T(x)$ and $g_1(x)$ are connected, at a given $x$. An analogous relation exists for $h_L(x)$, in which the Mellin moments of $h_L(x)$ can be split into a twist-2 and a twist-3 part. In the WW approximation, the twist three PDFs are fully determined by their twist-2 counterparts. Investigations for the WW relation for both $g_T(x)$ and $h_L(x)$ reveals that the approximation holds fora range of $x$, however violations are also observed. Besides twist-3 PDFs, twist-3 GPDs are also being pursed in lattice QCD~\cite{Bhattacharya:2021oyr}, is a totally unexplored research area experimentally.

\section{Interplay with Experimental Data}
\label{sec:LQCD-exp}

Phenomenological extractions of hadron structure from experimental data via the factorization theorem~\cite{Collins:1989gx} are subjected to an inverse problem, i.e., the inability to find a unique inverse function that solves quantities like PDFs in terms of the experimental data. The standard approach is the use of Bayesian inference by introducing suitable priors that allows us to regulate regions of parton's d.o.f that are not accessible by the data. In some cases such as in the GPDs sector, the inverse problem is more severe as their parton momentum fraction is effectively hidden in the data and it is not possible to constraint it in a unique manner. Due to these limitations, lattice QCD calculations of the PDFs and moments as alluded to in Sec.~\ref{parton_methods} provide additional constraints which have shown promising results in helping global analyses. Future lattice calculation will providing more constraints in this regard.  

In what follows we discuss specific cases where lattice QCD has the potential to expand our ability to delineate the quark and gluon structures of nucleons and nuclei:

\begin{itemize}

    \item Flavor separation of unpolarized and polarized PDFs is one the most challenging tasks in QCD global analysis, specially in the strange sector which plays an important role in precision electroweak physics, such as the determination of the $W$ mass, as well as for QCD model builders. The sensitivity of existing experiments to polarized and unpolarized strange PDF is rather weak even with the inclusion of Kaon production in $ep$ and $pp$ reactions and  $W+c$ data from LHC for the case of unpolarized PDFs. In this area, LQCD can make important contributions by providing extended correlation functions that are directly sensitive to the strange PDF. Similarly, calculations that are sensitive to $d/u$ at large-$x$ is of great importance to supporting the existing efforts at Jefferson Lab 12 program at high-$x$, which aim to understand nuclear effects in light nuclei~\cite{Cocuzza:2021rfn}.
 
 \item
 In the Euclidean path-integral formulation of the hadronic tensor~\cite{Liu:1993cv,Liu:1999ak} it was shown that there are connected and disconnected sea parton degrees of freedom along with the flavor dependence in the structure functions that can be accessed with lattice calculation. While experimental data are not formally sensitive to such separation, LQCD can provide new insights and constraints that can be integrated as part of QCD global analysis of hadron structure in general.          For instance, using LQCD calculation for moments of local operators,  an exploratory study was carried out in ~\cite{Hou:2021uoj}  where the input scale PDFs were parametrized in terms of the connected and disconnected sea components subjected to the constraints of LQCD moments. Their analysis found indications that the $\bar{d}-\bar{u}$ difference from the global analysis is associated with connected sea component in the path-integral formulation and the errors on the strange distribution can be reduced. In view of these findings, it is possible in the future to extend such analysis with LQCD computation of connected and disconnected sea components of non-local operators and incorporate them within the global analysis efforts and have a better understanding of the non-perturbative nature of hadron structure.

    \item Spin dependent PDFs such as helicity PDFs and transversity PDFs provide unique insights on the spin structure of nucleons. However since these quantities are experimentally accessible from spin asymmetries which are statistically bounded compared to unpolarized observables, the precision at which they can be inferred is limited relative to upolarized PDFs. LQCD can complement these efforts as shown in Ref.~\cite{Bringewatt:2020ixn} where the precision of existing lattice data provides constraints on helicity PDFs that are as competitive as the ones from experimental data. These efforts really demonstrate that, even at the current, early, stage of lattice calculations of non-local correlation functions, these can have a significant impact for the  understanding of the spin structure of hadrons. For instance, in a recent work~\cite{Zhou:2022wzm}, the current knowledge of gluon polarization was investigated finding that in the absence of positivity constraints, the sign of $\Delta g$ is still unknown using experimental data alone. Having direct LQCD calculations sensitive to gluon helicity will give the additional constraints that are needed to resolve the role of gluon polarization in the spin-puzzle.
    In addition, lattice calculations have identified recently that the smallness of the quark contribution to proton spin is due to the negative contribution in the DI corresponding  to  the  disconnected  pseudoscalar  density  and  topology  terms  in  the  anomalous Ward identity [172, 173].  Giving these findings,  it is a challenge for future global analysis to comprehend this aspect of the non-perturbative nature of the spin dependent PDFs.

    \item TMDs are a class of hadron structures with increasing complexity in terms of its theoretical definition and its connection with experimental observables. While tremendous theoretical efforts has been devoted in improving the theoretical aspects such as the TMD evolution, yet the full realization of a TMD QCD global analysis is still in its infancy relative to the state-of-the-art analysis of unpolarized collinear PDFs. Part of the difficulty relies on the ability to disentangle pertubative contributions from intrinsic non-perturbative ones which is by far more complicated that in the context of collinear physics. There are several approaches in the literature that addresses these challenges with many model parametrizations that aim to describe simultaneously all the available global data which are sensitive to TMDs. In this context, recent developments mentioned above from LQCD to compute fundamental quantities such as the CS kernel in the non-perturbative regime is an opportunity to guide TMD phenomenology. In this sense, a combined analysis of TMDs with lattice data will be ideal in order to confirm, for instance, the universality of the CS kernel across all high-energy reactions as well as the uniqueness of the non-perturbative hadronic components and their flavor dependence.   
    
    \item An important task for the future Jlab and EIC  program is to obtain reliable representations of GPDs using the experimental data on Compton form factors (CFFs). However, the majority of the planned experiments are insufficient to accomplish this task, because CFFs are two-dimensional functions, whereas GPDs are  three-dimensional ones. CFFs are convolutions of GPDs with complex kernels over a dimensionless kinematic variable~\cite{Ji:1996nm} where the  kernels are described by a controlled approximation in perturbative QCD. The GPD inverse problem can be in principle ill-posed if the estimator from the data extraction has a high variance. This problem can nevertheless be solved with a variety of theoretical constraints, particularly from lattice QCD. The first-principle lattice QCD calculations on GPDs are complementary in many ways. Lattice calculations, in fact, allow us to explore all of the kinematic dependence, including $\xi$, $t$, particularly the $x$ dependence that cannot be inferred directly from experimental data. On the other hand, they are not accurate in the end point regions $x=0, 1, \pm \xi$ unless one can study an extremely large-momentum nucleon,  something that is currently not possible. Therefore, only a complementary use of lattice computation and experimental data will allow  realistic three-dimensional images of the nucleon to be generated.

\end{itemize}

\section{Conclusions and outlook}
\label{sec:conclusions}

Since the previous Snowmass process, there have been tremendous breakthroughs in LQCD that enable us to compute from first principles the emergent quark and gluon structures inside nucleons and nuclei. These advances are opening new venues to access parton degrees of freedom encoded in quantities such as PDFs, TMDs and GPDs. In principle, such calculations have the potential to enable ab-initio predictions for hard processes in higher-energy reactions that can be compared directly with experimental data and provide a better understanding of QCD's emergent phenomena, as well as more reliable Standard-Model inputs to aid new-physics searches in many NP/HEP frontiers. 

However, currently, LQCD calculations are highly limited by the availability of computational resources and more are needed to achieve the ideal ab-initio predictions that can be tested experimentally. In this document, we have summarized the state of the art for such calculations and the associated computational challenges. We have also discussed the potential complementary and synergistic combinations with QCD global analysis using current precision LQCD calculations. Case studies found in spin physics show that even at this stage, some LQCD calculations are competitive with experimental data and allows us to carryout a \emph{hybrid} QCD global analysis, combining lattice and experimental data within a suitable Bayesian inference framework. Similar synergies can be also found in the context of meson structures relevant for future programs at AMBER at CERN and the tagged experiments at Jefferson Lab and the future  EIC. Moreover, for GPD studies, it is expected that LQCD inputs will become increasingly important to support programs for three-dimensional imaging of nucleons and nuclei at Jefferson Lab and the future EIC. 
With sufficient computational support, LQCD calculations can reach precision comparable with experimental data, which will greatly benefit the NP and HEP communities.

\begin{acknowledgments}
CJM, KNO, JWQ, DGR, and NS are supported in part by US DOE grant No. DE-AC05-06OR23177, under which Jefferson
Science Associates, LLC, manages and operates Jefferson Lab. 
PP supported by the U.S. Department of Energy, Office of through the Contract No. DE-SC0012704. 
The work of HL is partially supported by the US National Science Foundation under grant PHY 1653405 and by the  Research  Corporation  for  Science  Advancement through the Cottrell Scholar Award. The work of N.S. was supported by the DOE, Office of Science, Office of Nuclear Physics in the Early Career Program. The work of CPY is partially supported by the U.S.~National Science Foundation under Grant No.~PHY-2013791. JHZ is partially supported by National Natural Science Foundation of China under Grants No. 11975051 and 12061131006.
\end{acknowledgments}

\end{document}